\documentclass[conference]{IEEEtran}
\usepackage[]{acronym}
\usepackage{booktabs}
\usepackage{graphicx}
\usepackage{subfigure}
 \usepackage{lineno}

\ifCLASSINFOpdf
\else
\fi

\begin{document}
\bstctlcite{IEEEexample:BSTcontrol}

%
\title{Electric Vehicles Charging Control based on \\ 
Future Internet Generic Enablers}

\author{\IEEEauthorblockN{Andrea Lanna, Francesco Liberati, Letterio Zuccaro, Alessandro {Di Giorgio}}
\IEEEauthorblockA{Department of Computer, Control and Management Engineering ``Antonio Ruberti"\\
``Sapienza'' University of Rome\\
Via Ariosto, 25 - 00185 Rome, Italy\\
email: \textit{\{lanna,liberati,zuccaro,digiorgio\}}@diag.uniroma1.it}}



\maketitle

\acrodef{API}{Application Programming Interface}
\acrodef{CDI}{Connected Device Interface}
\acrodef{DSE}{Domain Specific Enabler}
\acrodef{DSO}{Distribution System Operator}
\acrodef{EMM}{Electric Mobility Management}
\acrodef{EV}{Electric Vehicle}
\acrodef{EVSE}{Electric Vehicle Supply Equipment}
\acrodef{GE}{Generic Enabler}
\acrodef{ICT}{Information and Communication Technology}
\acrodef{LAC}{Load Area Controller}
\acrodef{LA}{Load Area}
\acrodef{MLAA}{Macro Load Area Aggregator}
\acrodef{NetIC}{Network Information and Controller}
\acrodef{POD}{Points Of Delivery}
\acrodef{QoE}{Quality of Experience}
\acrodef{SGAM}{Smart Grid Architecture Model}
\acrodef{TSO}{Transmission System Operator}
\acrodef{UA}{Usage Area}

\begin{abstract}

In this paper a rationale for the deployment of Future Internet based applications in the field of \acp{EV} smart charging is presented. The focus is on the \ac{CDI} \ac{GE} and the \ac{NetIC} \ac{GE}, which are recognized to have a potential impact on the charging control problem and the configuration of communications networks within reconfigurable clusters of charging points. The CDI GE can be used for capturing the driver feedback in terms of \ac{QoE} in those situations where the charging power is abruptly limited as a consequence of short term grid needs, like the shedding action asked by the Transmission System Operator to the Distribution System Operator aimed at clearing networks contingencies due to the loss of a transmission line or large wind power fluctuations. The NetIC GE can be used when a master \ac{EVSE} hosts the Load Area Controller, responsible for managing simultaneous charging sessions within a given \ac{LA}; the reconfiguration of distribution grid topology results in shift of EVSEs among LAs, then reallocation of slave EVSEs is needed. Involved actors, equipment, communications and processes are identified through the standardized framework provided by the \ac{SGAM}. \par
\end{abstract}



%
\IEEEpeerreviewmaketitle

\acresetall
\section{Introduction}
In the last years, a relevant improvement of \ac{ICT} has taken place, producing smarter, smaller and faster devices which progressively require more and more connectivity. As a result, Internet based services are entering a new phase of mass deployment which is expected to raise a huge number of new opportunities but also new challenges in terms of scalability, capacity, throughput, mobility and trust. The European Union is addressing these challenges through a multidisciplinary approach leaded by strong European industrial stakeholders, supported by academics and innovative SMEs, in order to develop the devices, interfaces, networks and services required to support the future networked society and economy.
The Future Internet will contribute to close the gap between technology and applications and between the EU innovation and competitiveness \cite{GS_CFI13}. The European Future Internet ambitions include \cite{whPaFI}:\par
\begin{itemize}
\item providing the European citizens and industry with better and smarter services and applications that keep, extend in time, or enhance their quality of life and business;
\item fostering the creation of a new extended economy environment over the net, accessible by stakeholders in all member states, which guarantees service provision, service delivery, traceability, information quality;
\item leveraging the enlargement of service offering over the net, allowing a wider range of better quality-enabled services to all economy stakeholders.
\end{itemize}
In the last European Framework Programme (EU-FP7), \hbox{FI-WARE}\footnote{FI-WARE: Future Internet Core Platform, EU FP7 project n. 28524} and its follow-on FI-CORE\footnote{FI-CORE: Future Internet Core, EU FP7 project n. 632893} were the main financed projects in communication work-programme that represent the reference projects for Future Internet and its pioneer implementations \cite{HK_ICOIN12}. From these, others applicative projects have been launched in order to apply the conceptual solutions reached within FI-WARE/FI-CORE in specific scenarios (Health, Transport, Energy, \dots). FI-WARE/FI-CORE proposed a \textit {cognitive} architecture (\cite{MC_TFI11} and \cite{RC_IWC02}) in order to provide a truly open, public and royalty-free architecture, reduce obstacles and foster innovation and entrepreneurship. FI-WARE/FI-CORE are based upon elements called \acp{GE} which offer reusable and commonly shared functions serving a multiplicity of different sectors. In addition to \acp{GE}, in the FI-WARE/FI-CORE catalogue \acp{DSE} are also present which are enablers common to multiple applications but all of them specific to a very limited set of \acp{UA}. 
One of the key aspects of Future Internet is given by the introduction of \ac{QoE} concept, namely ``\textit{the overall acceptability of an application or service, as perceived subjectively by the end-user}'' \cite{FDP_FNMS12}. This fact implied a strong changing, from a model based on tangible objectives to a system leaded by personal perceptions. Another interesting point of view about the \ac{QoE} is given in \cite{CB_SJ14}, where it is defined as the perception that the user has about the performance of the service when he uses an application and about how this application is usable.\par

In this paper, a discussion about the role that the Future Internet can assume in the field of \acp{EV} smart charging is provided, focusing on the two \acp{GE} recognized to have a potential impact on the charging control problem and the configuration of communications networks within reconfigurable clusters of charging points, namely the \ac{CDI} and \ac{NetIC}. In Section 2, the GEs are described and their possible role in the considered smart charging scenario is provided. In Section 3 the impact on grid equipment and operation is clarified making use of the standardized \ac{SGAM}. 
In Section 4 the conclusions are drawn.
\bigskip

\section{Generic Enablers and proposed exploitation in the EV scenario}
\subsection{Connected Device Interface Generic Enabler - CDI GE}
The CDI GE provides the means to detect and to optimally exploit capabilities, resources and aspects about the
status of connected devices \cite{MCF_EEEIC13}, through the implementation of interfaces and \acp{API} towards device features.
In detail, the CDI GE provides three main interfaces:
\begin{itemize}
\item The On Device subsystem exposes a common set of \acp{API} to exploit native devices capabilities. Such capabilities include device sensors (camera, mic, geo-location, device orientation), device information (CPU, disk space), personal data services (Contact, Calendar, Gallery), messaging (SMS, MMS, email). This subsystem contains also models and primitives for the assessment of the \ac{QoE} level of the user.
\item The Remote Management subsystem presents an externally accessible interface which supports external services for device management and configuration.
\item The Mobility Manager connects to the Service Capability Connectivity \& Control (S3C) Generic Enabler, a
system in charge of configuring network policies, routing rules to fulfil QoS requirements.
\end{itemize}
The CDI GE implementation is available to users and developers in the FI-WARE/FI-CORE Catalogue, under the item ``A-CDI''. The CDI API are integrated in the Webinos platform and developed in JavaScript, ready to be used in HTML5 applications.
\par\smallskip
The following CDI \textit{On Device} functionalities can be exploited:
\begin{itemize}
\item Detection of user position (\textit{Device Sensors/Geolocation API}). When the user sends a Charging Request to the control system, the information about his position could help in choosing the more suitable EVSE (i.e. the nearest one) to satisfy the request.
\item Sending notification to the user (\textit{Messaging API}). When a load shedding event occurs, it may be helpful to notify the user about possible changes/reschedules concerning his charging session.
\item Retrieving user feedback (\textit{QoE API}). After a load shedding event, the degree of satisfaction/acceptance about changes/reschedules can be used to influence the \ac{LAC} process of charging control.
\end{itemize}
By exploiting the above-mentioned CDI GE functionalities, the \ac{LAC} algorithm for scheduling the charging (see for example \cite{ADG_MED12,ADG_MED13,ADG_CDC13} and \cite{ADG_CEP14}), at the core of the EVSE DSE functionalities, can be upgraded with the direct inclusion of the feedback from the EV users (gathered via the CDI).  Based on this feedback, a load shedding algorithm preserving the quality of the charging service provided to the users can be designed to enable the \ac{DSO} to provide the \ac{TSO} with enhanced load shedding services.\par
\begin{figure}[h] 
   \centering
   \includegraphics[width=3 in]{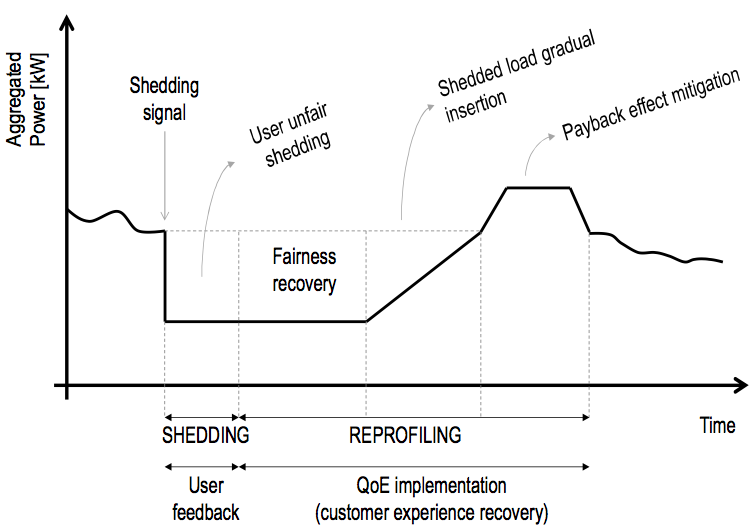} 
   \caption{Extended Load shedding process}
   \label{LAC1}
\end{figure}

\begin{figure}[h!] 
   \centering
   \includegraphics[width=3 in]{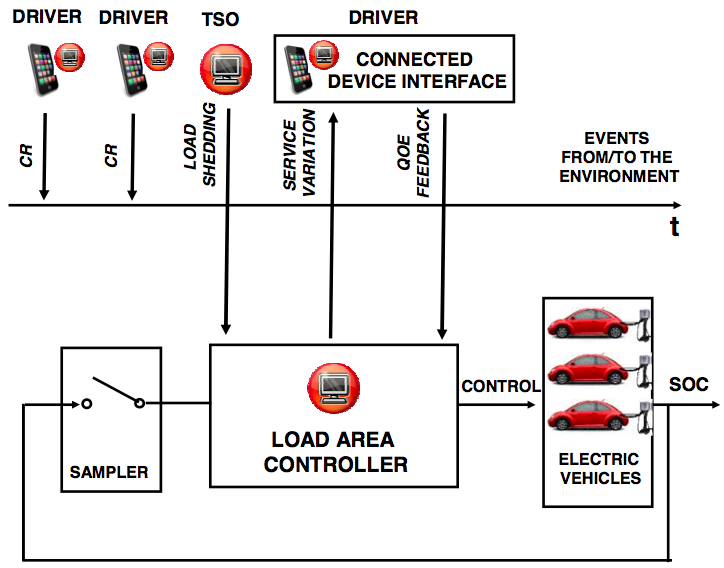} 
   \caption{LAC working logic}
   \label{LAC2}
\end{figure}

The implementation of load shedding plans accepted by the DSO can be executed on public charging stations managed by the \ac{EMM} system. In principle, EMM system, as a system operating charging stations, is owned by a dedicated \ac{EVSE} Operator. In Italy, upon request of national Authority of Electricity and Gas, a specific business model is tested, by which the charging stations are owned and managed by the DSO.
This means that, once load shedding plans are traded with TSO, the DSO can implement it using the functionalities and algorithms described and possibly embedded in the EVSE Operator system, running a load optimization program spread over the chosen \ac{POD}. Such a load optimization program, which also fulfill other use cases beyond load shedding (\cite{ADG_CEP14} and \cite{FL_MED14}), could be in principle running as a separate entity. \par
This implies that the DSO could be able to generate value modulating e-mobility load at Low Voltage and Medium Voltage level. The value is generated by aggregating power flexibility from customers and trading it against the needs of TSO at High Voltage level. Such a conversion could be managed by the DSO over a wide area (possibly nationwide), without producing impacts at LV and MV level, but generating a positive or negative peak in power request capable of meeting balancing needs of TSO \cite{JM_TSG10}. As such trading should be flexible and fast enough to be competitive against switching on and off of conventional power plants, it should be designed to be run within a matter of seconds. This means that the customer interaction, with his allowance for the proposed reduction of degree of freedom, happens at a second step: the DSO implements load shedding plan, i.e. either delaying some charging processes or lowering their power output; subsequently, the customers are prompted with a notification based on CDI generic enabler that harvest customers satisfaction for the charge management plan, and EMM implements corrective actions through a dedicated EVSE DSE algorithm considering the feedbacks of users providing an answer. The phases characterizing the proposed load shedding process and the related LAC working logic are reported in Fig. \ref{LAC1} and Fig. \ref{LAC2}. 

\subsection{Network Information and Controller Generic Enabler - NetIC GE}
The NetIC GE defines a common and unified interface that allows utilizing network resources as a service. NetIC not only abstracts the physical network resources, such as nodes, boards, ports and links that connect them, into URI, but it also provides an abstraction of operations that can be used to control and manipulate them in a RESTful fashion. It exposes network status information and it enables a certain level of programmability within the network, e.g., concerning flow processing, routing, addressing, and resource management at flow and circuit level. In this way, different parties such as network providers, network operator and virtual network providers can simultaneously utilize the network infrastructure. One of the advantages of ÒNetwork as a ServiceÓ paradigm is the possibility for different network operators to implement tailored network mechanisms even relying on the same common network infrastructure (own addressing schemes, protocols, dynamic tunnels, etc.) \cite{MRN_ICFIT11}.\par
Potential users of NetIC interfaces include network service providers or other components of FI-WARE/FI-CORE, such as cloud hosting. Network operators, but also virtual network operators and service providers may access (within the constraints defined by contracts with the operators) the open networks to both set control policies and optimally exploit the network capabilities, and also to retrieve information and statistics about network utilization. For example, a typical use case scenario might be a smart application where the \ac{DSO} which manages its own smart grid, needs a dedicated virtual network. The advantages of NetIC GE is that the network management and configuration effort is minimal, this GE improves the ability to control efficiently, and manage resources in complex infrastructures and to scale up management in those scenarios.\par\smallskip
\begin{figure}[h] 
   \centering
   \includegraphics[width=3 in]{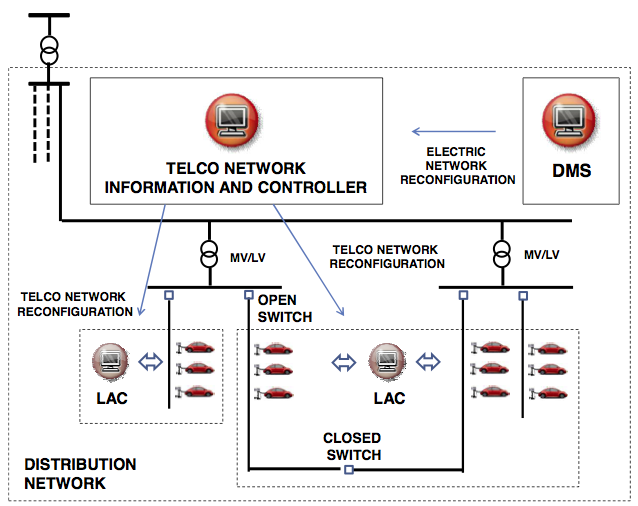} 
   \caption{Network reconfiguration scenario.}
   \label{NetRec}
\end{figure}
\begin{figure*}[t!]
\centering
\includegraphics[width=7.2in]{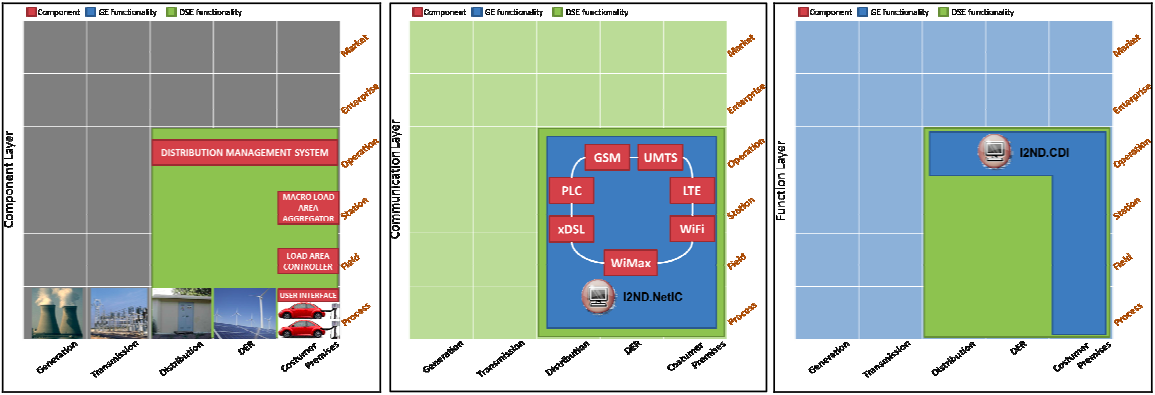}
\caption{Component (left), Communication (center) and Functional (right) SGAM layers. }
\label{fig:sgam}
\end{figure*}
The NetIC functionalities can be exploited in order to provide the DSE control system with the reconfiguration capabilities needed to react to grid topologies changes operated by the DSO through its Distribution Management/SCADA System  \cite{AM_TPD09} which, on a daily basis, performs adjustments in the grid topology (e.g. by acting on tie switches, see Fig. \ref{NetRec}) in order to maintain a convenient network configuration. In that case, a set of PODs is shifted from the control of a LAC to another one and the DSE has to maintain (thanks to the NetIC functionalities) proper connectivity among the \acp{MLAA}, as defined in \cite{FL_MED14}, the LACs and all the charging stations under control. The NetIC GE enables the communication network programmability: so, network-aware applications built on top of the NetIC API can handle the following issues:
\begin{itemize}
\item Grid reconfiguration: due to fault events or load balancing problems the DSO might need to re-locate the PODs from a LA to another (i.e. \cite{OptPla} and \cite{SC_MED12} show some of the electrical and the communication problems arising from a fault). The connection between the related LACs and the PODs is guaranteed in a seamless way by the NetIC GE.
\item Network resources optimization: the DSO might use low bandwidth/throughput network to connect the Electric Meter to the servers for electronic billing. The management of limited and valuable resources can be efficiently optimized in order to guarantee a certain level of QoS in the virtual network paths. The NetIC GE provides scalable network management APIs which enable resource virtualization and isolation.
\item Network policy: firewall rules or load balancing routing algorithms can be deployed with the ease of writing a single program with no need to deploy it in all the nodes \cite{GO_MED13}. In fact, in the open networking paradigm all the control plane functionalities (the intelligence) resides in the Network Controller, whilst the network nodes just forward packets.
\end{itemize}
\subsection{Mapping onto the SGAM}
Making reference to the SGAM\footnote{SGAM (Smart Grid Architecture Model) is defined by CEN-CENELEC-ETSI SG-CG and available at http://www.cencenelec.eu}, Fig. \ref{fig:sgam} shows the mapping of the two GEs onto the SGAM. These GEs work on the Component, Communication and Functional interoperability layers and, within each layer, on the zones from Process to Operation and the domains from Customer to Distribution. Fig. \ref{fig:sgam} shows also how the NetIC and CDI functionalities are distributed on the two different interoperability layers but in the same zones-domains area. \section{Conclusions}
In this paper the possible role of Future Internet functionalities in the field of \acp{EV} smart charging has been discussed, highlighting the opportunity for significant advancements when joining competencies and innovations from both research domains. A first improvement can be identified in the integration of the \acp{GE} as follows \cite{castrucci2011key}:
\begin{itemize}
\item The application of the device functionalities provided by the CDI GE is itself an added value in the smart energy context. In particular, its integration may facilitate the interaction with the user/driver, which could take a more active role. In addition, the DSO may benefit of that, in the perspective of improving the offered services and enhancing the customer experience.
\item The NetIC integration in a Smart Grid scenario is a significant step towards the Future Internet Core Platform integration. The need for tailored network policies on routing, security, load balancing, reconfiguration, fault tolerance, etc. in the DSO network, is fulfilled by the utilization of the NetIC GE. Moreover the policy rules are maintainable and efficient, since they are deployed on a single entity, the application built on top of the NetIC API \cite{AP_FNMS13}.
\end{itemize}
As far as concerns charging control capabilities, a control strategy based on those two GEs go beyond the state of the art particularly by providing the DSO with the possibility to control electromobility load profiles, taking into account both network constraints/requirements and the feedbacks from the EV users, in order to achieve efficient operation of the grid while improving the perceived quality of the charging services. As a matter of fact future works will be dedicated to the mathematical extension of the charging control framework presented in \cite{MCF_EPSR11}, \cite{MCF_SPD14} and \cite{DS_EPSR14}, by integrating user feedbacks for enabling the load shedding use case and the local matching of charging load with the power generated by partially controllable wind turbines \cite{ADG_MED10}.

\section*{Acknowledgment}

This work is partially funded by the FI-CORE (grant agreement no. 632893) and MOBINCITY (grant agreement no. 314328) projects, funded by the European Commission under the 7th Framework Programme.\\
The authors would like to thank Prof. F. Delli Priscoli, Prof. R. Cusani, Prof. S. Battilotti and Dr. A. Pietrabissa for the provisioning of relevant data and the helpful suggestions.

\IEEEtriggeratref{18}



\bibliographystyle{myIEEEtran}
\bibliography{IEEEabrv}

\begin{thebibliography}{10}
\providecommand{\url}[1]{#1}
\csname url@samestyle\endcsname
\providecommand{\newblock}{\relax}
\providecommand{\bibinfo}[2]{#2}
\providecommand{\BIBentrySTDinterwordspacing}{\spaceskip=0pt\relax}
\providecommand{\BIBentryALTinterwordstretchfactor}{4}
\providecommand{\BIBentryALTinterwordspacing}{\spaceskip=\fontdimen2\font plus
\BIBentryALTinterwordstretchfactor\fontdimen3\font minus
  \fontdimen4\font\relax}
\providecommand{\BIBforeignlanguage}[2]{{%
\expandafter\ifx\csname l@#1\endcsname\relax
\typeout{** WARNING: IEEEtran.bst: No hyphenation pattern has been}%
\typeout{** loaded for the language `#1'. Using the pattern for}%
\typeout{** the default language instead.}%
\else
\language=\csname l@#1\endcsname
\fi
#2}}
\providecommand{\BIBdecl}{\relax}
\BIBdecl

\bibitem{GS_CFI13}
G.~Sallai, ``{Chapters of Future Internet research},'' in \emph{Cognitive
  Infocommunications (CogInfoCom), 2013 IEEE 4th International Conference on},
  pp. 161--166, Dec 2013. DOI: 10.1109/CogInfoCom.2013.6719233

\bibitem{whPaFI}
\BIBentryALTinterwordspacing
{The European Future Internet Initiative}, ``{White paper on the Future
  Internet PPP definition},'' 2010. [Online]. Available:
  \url{www.future-internet.eu}
\BIBentrySTDinterwordspacing

\bibitem{HK_ICOIN12}
H.~Kim and S.~Lee, ``{FiRST Cloud Aggregate Manager development over FiRST:
  Future Internet testbed},'' in \emph{Information Networking (ICOIN), 2012
  International Conference on}, pp. 539--544, Feb 2012. DOI:
  10.1109/ICOIN.2012.6164436

\bibitem{MC_TFI11}
M.~Castrucci, F.~Delli~Priscoli, A.~Pietrabissa, and V.~Suraci,
  ``\BIBforeignlanguage{English}{A cognitive {Future Internet} architecture},''
  in \emph{\BIBforeignlanguage{English}{The Future Internet}}, ser. Lecture
  Notes in Computer Science.\hskip 1em plus 0.5em minus 0.4em\relax Springer
  Berlin Heidelberg, 2011, vol. 6656, pp. 91--102. DOI:
  10.1007/978-3-642-20898-0\_7

\bibitem{RC_IWC02}
R.~Cusani, F.~Delli~Priscoli, G.~Ferrari, and M.~Torregiani, ``{A novel MAC and
  scheduling strategy to guarantee QoS for-the new-generation WIND-FLEX
  wireless LAN},'' \emph{Wireless Communications, IEEE}, vol.~9, no.~3, pp.
  46--56, June 2002. DOI: 10.1109/MWC.2002.1016711

\bibitem{FDP_FNMS12}
F.~Delli~Priscoli, V.~Suraci, A.~Pietrabissa, and M.~Iannone, ``{Modelling
  Quality of Experience in Future Internet networks},'' in \emph{Future Network
  Mobile Summit (FutureNetw), 2012}, pp. 1--9, July 2012. ISBN
  978-1-4673-0320-0

\bibitem{CB_SJ14}
C.~Bruni, F.~Delli~Priscoli, G.~Koch, A.~Palo, and A.~Pietrabissa, ``{Quality
  of Experience provision in the Future Internet},'' \emph{Systems Journal,
  IEEE}, vol.~PP, no.~99, pp. 1--11, 2014. DOI: 10.1109/JSYST.2014.2344658

\bibitem{MCF_EEEIC13}
M.~Falvo, L.~Martirano, D.~Sbordone, and E.~Bocci, ``Technologies for smart
  grids: A brief review,'' in \emph{Environment and Electrical Engineering
  (EEEIC), 2013 12th International Conference on}, pp. 369--375, May 2013. DOI:
  10.1109/EEEIC.2013.6549544

\bibitem{ADG_MED12}
A.~Di~Giorgio, F.~Liberati, and S.~Canale, ``Optimal electric vehicles to grid
  power control for active demand services in distribution grids,'' in
  \emph{Control Automation (MED), 2012 20th Mediterranean Conference on}, pp.
  1309--1315, July 2012. DOI: 10.1109/MED.2012.6265820

\bibitem{ADG_MED13}
A.~Di~Giorgio, F.~Liberati, and S.~Canale, ``{IEC} 61851 compliant electric
  vehicle charging control in smartgrids,'' in \emph{Control Automation (MED),
  2013 21st Mediterranean Conference on}, pp. 1329--1335, June 2013. DOI:
  10.1109/MED.2013.6608892

\bibitem{ADG_CDC13}
A.~Di~Giorgio, F.~Liberati, and A.~Pietrabissa, ``On-board stochastic control
  of electric vehicle recharging,'' in \emph{Decision and Control (CDC), 2013
  IEEE 52nd Annual Conference on}, pp. 5710--5715, Dec 2013. DOI:
  10.1109/CDC.2013.6760789

\bibitem{ADG_CEP14}
A.~Di~Giorgio, F.~Liberati, and S.~Canale, ``Electric vehicles charging control
  in a smart grid: A model predictive control approach,'' \emph{Control
  Engineering Practice}, vol.~22, pp. 147 -- 162, 2014. DOI:
  10.1016/j.conengprac.2013.10.005

\bibitem{FL_MED14}
F.~Liberati, A.~Mercurio, L.~Zuccaro, A.~Tortorelli, and A.~Di~Giorgio,
  ``Electric vehicles charging load reprofiling,'' in \emph{Control Automation
  (MED), 2014 22st Mediterranean Conference on}, June 2014.

\bibitem{JM_TSG10}
J.~Medina, N.~Muller, and I.~Roytelman, ``Demand response and distribution grid
  operations: Opportunities and challenges,'' \emph{Smart Grid, IEEE
  Transactions on}, vol.~1, no.~2, pp. 193--198, Sept 2010. DOI:
  10.1109/TSG.2010.2050156

\bibitem{MRN_ICFIT11}
M.~R. Nascimento, C.~E. Rothenberg, M.~R. Salvador, C.~N.~A. Corr\^{e}a, S.~C.
  de~Lucena, and M.~F. Magalh\~{a}es, ``Virtual routers as a service: The
  routeflow approach leveraging software-defined networks,'' in
  \emph{Proceedings of the 6th International Conference on Future Internet
  Technologies}, ser. CFI '11, pp. 34--37.\hskip 1em plus 0.5em minus
  0.4em\relax New York, NY, USA: ACM, 2011. DOI: 10.1145/2002396.2002405

\bibitem{AM_TPD09}
A.~Mercurio, A.~Di~Giorgio, and P.~Cioci, ``Open-source implementation of
  monitoring and controlling services for {EMS/SCADA} systems by means of web
  services - {IEC 61850} and {IEC 61970} standards,'' \emph{Power Delivery,
  IEEE Transactions on}, vol.~24, no.~3, pp. 1148--1153, July 2009. DOI:
  10.1109/TPWRD.2008.2008461

\bibitem{OptPla}
S.~Canale, A.~Di~Giorgio, A.~Lanna, A.~Mercurio, M.~Panfili, and
  A.~Pietrabissa, ``Optimal planning and routing in medium voltage powerline
  communications networks,'' \emph{Smart Grid, IEEE Transactions on}, vol.~4,
  no.~2, pp. 711--719, June 2013. DOI: 10.1109/TSG.2012.2212469

\bibitem{SC_MED12}
S.~Canale, F.~Delli~Priscoli, A.~Di~Giorgio, A.~Lanna, A.~Mercurio, M.~Panfili,
  and A.~Pietrabissa, ``Resilient planning of powerline communications networks
  over medium voltage distribution grids,'' in \emph{Control Automation (MED),
  2012 20th Mediterranean Conference on}, pp. 710--715, July 2012. DOI:
  10.1109/MED.2012.6265721

\bibitem{GO_MED13}
G.~Oddi and A.~Pietrabissa, ``A distributed multi-path algorithm for wireless
  ad-hoc networks based on wardrop routing,'' in \emph{Control Automation
  (MED), 2013 21st Mediterranean Conference on}, pp. 930--935, June 2013. DOI:
  10.1109/MED.2013.6608833

\bibitem{castrucci2011key}
M.~Castrucci, M.~Cecchi, F.~{Delli Priscoli}, L.~Fogliati, P.~Garino, and
  V.~Suraci, ``{Key concepts for the Future Internet architecture},'' in
  \emph{Future Network \& Mobile Summit (FutureNetw), 2011}, pp. 1--10.\hskip
  1em plus 0.5em minus 0.4em\relax IEEE, 2011.

\bibitem{AP_FNMS13}
A.~Palo, L.~Zuccaro, A.~Simeoni, V.~Suraci, L.~Musto, and P.~Garino, ``A common
  open interface to programmatically control and supervise open networks in the
  future internet,'' in \emph{Future Network and Mobile Summit
  (FutureNetworkSummit), 2013}, pp. 1--9, July 2013.

\bibitem{MCF_EPSR11}
M.~C. Falvo, R.~Lamedica, R.~Bartoni, and G.~Maranzano, ``Energy management in
  metro-transit systems: An innovative proposal toward an integrated and
  sustainable urban mobility system including plug-in electric vehicles,''
  \emph{Electric Power Systems Research}, vol.~81, no.~12, pp. 2127 -- 2138,
  2011. DOI: 10.1016/j.epsr.2011.08.004

\bibitem{MCF_SPD14}
M.~C. Falvo, G.~Graditi, and P.~Siano, ``Electric vehicles integration in
  demand response programs,'' in \emph{Power Electronics, Electrical Drives,
  Automation and Motion (SPEEDAM), 2014 International Symposium on}, pp.
  548--553, June 2014. DOI: 10.1109/SPEEDAM.2014.6872126

\bibitem{DS_EPSR14}
D.~Sbordone, I.~Bertini, B.~D. Pietra, M.~Falvo, A.~Genovese, and L.~Martirano,
  ``{EV} fast charging stations and energy storage technologies: A real
  implementation in the smart micro grid paradigm,'' \emph{Electric Power
  Systems Research}, 2014. DOI: 10.1016/j.epsr.2014.07.033

\bibitem{ADG_MED10}
A.~Di~Giorgio, L.~Pimpinella, and A.~Mercurio, ``A feedback linearization based
  wind turbine control system for ancillary services and standard steady state
  operation,'' in \emph{Control Automation (MED), 2010 18th Mediterranean
  Conference on}, pp. 1585--1590, June 2010. DOI: 10.1109/MED.2010.5547821

\end{thebibliography}
%

\end{document}